\documentclass [12pt]{article}
\usepackage{amssymb,amsmath}
\usepackage{graphicx}
\bigskip

\begin{document}

\author{Z.D. Genchev \thanks{Inst Electron B.A.S., 1784 Sofia, Bulgaria. E-mail \tt{zgenchev@ie.bas.bg.}} \and T.L. Boyadjiev\thanks{Faculty of Mathematics and Computer Sciences, University of Sofia ``St. Kliment Ohridski",1163 Sofia, Bulgaria. E-mail \tt{todorlb@fmi.uni-sofia.bg}}}

\title{On the solution of the modified Ginzburg-Landau type equation for one-dimensional superconductor in presence of a normal layer}

\date{}

\maketitle

\begin{abstract}

We perform an analytical and numerical study of the crossover from the Josephson effect to the bulk superconducting flow for two identical one-dimensional superconductors, co-existing with a layer of normal material.

A generalized Ginzburg-Landau (GL) model, proposed by S.J. Chapman, Q. Du and M.D. Gunzburger \cite{chdugun} was used in modeling the whole structure. When the thickness of the normal layer is very small, the introduction of three effective $\delta$-function potentials of specified strength leads to an exact analytical solution of the modified stationary GL equation.

The resulting current density-phase offset relation is analyzed numerically. We show that the critical Josephson current density $j_c$ corresponds to a bifurcation of the solutions of the nonlinear boundary value problem coupled with the modified GL-equation. The influence of the second term in the Fourier-decomposition of the supercurrent density-phase relation is also investigated.

We derive also a simple analytical formula for the critical Josephson current.

\end{abstract}

\section{Introduction}

If two superconductors are weakly coupled and have a phase difference $\Delta \phi $ that is not $0$ or $\pi$, a zero-voltage supercurrent flows from one to the other usually at a rate proportional to $\sin \Delta \phi $, as predicted by Josephson \cite{J1962} in 1962 for the case of junction consisting of an insulating oxide layer between two identical superconductors. A DC supercurrent can flow through such junction in the absence of a voltage difference between both superconductors in such a way that
\begin{equation}
j_{s}\left( \Delta \phi \right) =j_{c} \sin \Delta \phi .
\label{F1}
\end{equation}

After the discovery of the Josephson effect, it became clear that, apart from an insulating tunnel structure (SIS), any sufficiently short localized weak link such as a very short constriction in the cross-section of a superconductor, a point contact between two superconductors, as well as two superconductors separated by a thin layer of normal metal, could be used as a Josephson junction, obeying the current-phase relations, usually different from (\ref{F1}). This fact forced Licharev \cite{Licharev} and Waldram \cite{Waldram} to dwell upon a broad minded definition: a weak link is supposed to show a Josephson behavior if the supercurrent-phase relation is a single-valued and analytical function, which can be represented as a Fourier series
\begin{equation}
i_{s}\left( \Delta \phi \right) =\sum_{n=1}^{\infty }a_{n}\sin
\left( n\Delta \phi \right).  \label{B}
\end{equation}

The crossover between an ideal Josephson behavior and an uniform superconducting flow was studied by solving exactly the usual Ginzburg-Landau (GL) equation for a 1-D superconductor in the presence of an effective $\delta $-function potential of arbitrary strength (see, for example \cite{Sols}). Recently, a modified GL type model has been formulated \cite {chdugun}. This model could equally well be applied to a boundary between different superconductors, superconductor-insulator, superconductor-normal metal. The purpose of our paper is to apply this modified GL model for calculating the supercurrent-phase relation and the crossover between Josephson behavior and uniform superconducting flow.

\section{Statement of the Problem}

The bifurcation analysis of superconducting solutions from the normal solutions for the 1-D GL equations in the presence of an external magnetic field was considered in \cite{aftchap}. Here, we also consider the 1-D case, but on the basis of the generalized (modified) GL model \cite{chdugun}. 

By neglecting the external and the self-induced magnetic fields we have
\noindent in the superconducting domain $x \in (-\infty, \> -\tilde{d}/2)$ and $x \in (\tilde{d}/2, \> \infty)$
\begin{equation}
-\frac{\rlap{\protect\rule[1.1ex]{.325em}{.1ex}}h%
^{2}}{2m_{s}}\widetilde{\psi }^{\prime \prime }(x)+a_{s}\widetilde{\psi }%
(x)+b_{s}|\widetilde{\psi }|^{2}\widetilde{\psi }(x)=0,
\label{eq1}
\end{equation}
$$j_{s}=-\frac{i\hbar e_{s}}{2m_{s}}\left[ \widetilde{\psi }^{*}\widetilde{%
\psi }^{\prime }-\widetilde{\psi }\widetilde{\psi }^{*\prime}\right],$$
\noindent in the normal domain $x \in (-d/2, \> d/2)$
\begin{equation}
-\frac{\hbar ^{2}}{2m_{n}}\widetilde{\psi }^{\prime \prime }(x)+a_{n}%
\widetilde{\psi }(x)+b_{n}|\widetilde{\psi }|^{2}\widetilde{\psi
}(x)=0, \label{eq3}
\end{equation}
$$j_{n}=-\frac{i\hbar e_{n}}{2m_{n}}\left[ \widetilde{\psi }^{*}\widetilde{%
\psi }^{\prime }-\widetilde{\psi }\widetilde{\psi }^{*\prime}\right], $$
\noindent and the boundary conditions at the N-S interfaces $x=\pm \> \tilde{d}/2$ are
$$\left[ \widetilde{\psi }\right] =0, \quad \left[ \frac{m_s}{m_n}\frac{d\widetilde{\psi }}{dx}\right] =0,$$
\noindent so that $j_s=j_n=j$. Here $[f]$ denotes, as usually \cite{chdugun}, the jump of the enclosed function $f(x)$ across the points $x=\pm \> \tilde{d}/2$. The quantities $e_{s}$ and $e_{n}$ are equal to the charge of the superconducting charge carriers, and in the following are both equal to twice the electron charge. We are free to arbitrary choose one of the masses $m_s$ and $m_n$. Usually $m_s$ is chosen to be twice the electron mass. This leaves $m_n$ as a parameter depending on the normal material.

We suppose that $a_{s}=-|a_{s}|,$ $a_{n}>0,$ $b_{s}>0,$ $b_{n}\geq 0$ and we define the coherence length
\begin{equation}
\xi =\frac{\hbar }{\sqrt{2m_{s}|a_{s}|}},  \label{eq7}
\end{equation}
\noindent as well as the dimensionless distances $z=x/ \xi,\quad d=\widetilde{d}/\xi$, the order parameter $\psi (z) = \widetilde{\psi}(z\xi) \sqrt{b_{s}/|a_{s}|}$, and the current density
\begin{equation}
J=\sqrt{\frac{m_{s}}{2|a_{s}|}}\frac{b_{s}}{|a_{s}|e_{s}}j_{s}.
\label{eq8c}
\end{equation}

In order to make further comparisons with other papers we introduce an equivalent representation of equation (\ref{eq8c})
$$J = \frac{2\pi \xi }{\phi _{0}}\Lambda j_{s},$$
\noindent where $\Lambda = m_{s}/n_{0}e_{s}^{2}=\mu _{0}\lambda ^{2}_{L}$, and $n_{0}=\sqrt{|a_{s}|/b_s}$ is the equilibrium concentration of superelectrons, $\lambda_L $ is the well-known London penetration depth, $\phi_{0}=h/2e=h/e_s$ is the magnetic flux quantum.

With the definitions given in (\ref{eq7})-(\ref{eq8c}) our problem is stated as follows
\begin{equation} \label{eq9}
\psi ^{\prime \prime }+\left( 1-|\psi |^{2}\right) \psi =0, \quad (-\infty <z<-d/2) \cup (d/2<z<\infty),
\end{equation}
$$J=-\frac{i}{2}\left[ \psi ^{*}\psi ^{\prime }-\psi \psi ^{*\prime }\right],$$
\begin{equation} \label{eq10}
\frac{m_{s}}{m_{n}}\psi ^{\prime \prime }-\frac{a_{n}}{|a_{s}|}\psi -\frac{b_{n}}{b_{s}}|\psi |^{2}\psi =0, \quad  |z|< d/2. 
\end{equation}
$$J=-\frac{i}{2}\frac{m_{s}}{m_{n}}\left[ \psi ^{*}\psi ^{\prime }-\psi \psi^{*\prime }\right],$$

\section{Analytical solution for a thin normal layer}

Let us introduce the parameters
$$m =\frac{m_{n}}{m_{s}}, \> \alpha=\frac{m_{n}a_{n}}{m_{s}|a_{s}|}=m \frac{a_n}{a_s}, \> \beta =\frac{m_{n}b_{n}}{m_{s}b_{s}}=m \frac{b_n}{b_s}.$$

In both Eqns. (\ref{eq9}), (\ref{eq10}) we set $\psi(z)=R(z)\exp \left[ i\varphi (z)\right]$ and we find
\begin{equation}
R^2(z)\varphi ^{\prime }(z)=\left\{
\begin{array}{l}
 m \> J,\quad |z|<d/2; \\
J,\qquad |z|>d/2,
\end{array}
\right.   \label{eq12}
\end{equation}
\begin{equation}
R^{\prime \prime }+R-R^{3}- \frac{J^2}{R^3}=0,\quad |z|>d/2,
\label{eq14}
\end{equation}
\begin{equation}
R^{\prime \prime } - \alpha R - \beta R^{3}- m^2 \frac{J^2}{R^3}=0,\quad |z|< d/2. \label{eq15}
\end{equation}

It is not surprising that the case $\alpha =-1,\> \beta =m =1$ corresponds to an uniform superconductor occupying the whole space ($-\infty, \infty$).

Let us introduce the function
\[
\delta \left( z;1-c\right) =1+\left( 1-c\right) \left[ \theta \left( z-\frac{%
d}{2}\right) -\theta \left( z+\frac{d}{2}\right) \right] =\left\{
\begin{array}{l}
1,\quad |z|>d/2; \\
c,\quad |z|<d/2.
\end{array}
\right.
\]

Here, $\theta (z)$ is the Heaviside function. Then Eqns. (\ref{eq14}), (\ref {eq15}) can be written in the whole space $(-\infty <z<\infty )$ as
\begin{equation}
R^{\prime \prime } + \delta \left( z;1+\alpha \right) R- \delta \left( z;1-\beta \right) R^3 - \delta \left(z;1-m ^{2}\right) \frac{J^2}{R^3}=0. \label{gench}
\end{equation}

If the thickness $d\to +0$, we have $\theta \left( z-\frac{d}{2}\right) -\theta \left( z+\frac{d}{2}\right) =-d\delta \left( z\right) ,$ so that the case of small $d$ can be formulated as follows
\begin{equation}
R^{\prime \prime }+\left[ 1-g_{1}\delta (z)\right] R - \left[ 1-g_{2}\delta \left( z\right) \right] R^3 - \left[ 1-g_{3}\delta \left( z\right) \right] \frac{J^2}{R^3}=0, \label{eq19}
\end{equation}
\noindent where we merely substitute $g_{1}\equiv \left( 1+\alpha \right) d,\> g_{2}\equiv \left(1-\beta \right) d,\> g_{3}\equiv \left( 1-m ^{2}\right) d.$

The solution of Eqn. (\ref{eq19} ) is found to be
\begin{equation}\label{eq21}
  R^2(z) = a + b \tanh^2[u(|z|+z_0)]
\end{equation}
\noindent where $ a(2-a)^2=8J^2$, $ 0\leq a \leq{2/3}$, $ b = 1-3 a/2$, $ u = \sqrt{b/2}$, $b=a B^2$, and the quantity $y_0 \equiv \tanh(u z_0)$ satisfies the following equation ($ 0\leq y_0\leq 1 $)
\begin{equation} \label{f30}
  \sqrt{2 b} B^2 y_0 (1-y^2_0) = g_1(1+B^2 y^2_0) - a
  g_2 (1+B^2y^2_0)^2 - \frac{g_3 J^2}{a^2 (1+B^2 y^2_0)}.
\end{equation}

If $g_2=0$ and $g_3=0$ we recover Eqn. (15) from \cite{Sols}.

Now we will introduce the phase offset $\Delta \varphi$. Eqn. (\ref{eq12}) can be rewritten as
$$\varphi^\prime = \frac {J}{R^2(z)} \> \delta(z;1-m)],$$
\noindent so that for small $d$ we have
\begin{equation}\label{eq25}
  \varphi^\prime = \frac {J}{R^2(z)} \> [1-d(1-m)\,\delta(z)]
\end{equation}

Due to the fact that the boundary conditions for the order parameter at $z\rightarrow\pm\infty$ must be
$$ R^{\prime}(\pm\infty)=0, \quad \varphi(z)= \frac{J}{R^2_{\infty}}z \pm \frac{\Delta\varphi}{2}, \> z\rightarrow\pm\infty,$$
\noindent we derive from Eqn. (\ref{eq25})
\begin{equation}\label{eq27}
  \Delta\varphi=-\frac{J
  d(1-m)}{R^2(0)}+J\int_{-\infty}^{\infty}(\frac{1}{R^2(z)}-\frac{1}{R^2_{\infty}})dz.
\end{equation}
\noindent Here $ R_{\infty} = R(\pm\infty) > 0$ and $ R^2(0)=a+b y^2_0=a(1+B^2 y^2_0)$. Then by calculating the integral in Eqn. (\ref{eq27}), we finally find
\begin{equation}\label{eq28}
 \Delta\varphi=2\{\arctan B-\arctan(B y_0)\}-\frac{d
 J(1-m)}{a(1+B^2y^2_0)}.
\end{equation}
\noindent If $m = 1$ this result formally coincides with Eqn. (16) in \cite{Sols}.

\section{Numerical Results}

The Generalized Continuous Analogue of Newton's Method (see the survey by Puzynin et al. \cite{puzynin}) for solving the nonlinear ODE (\ref{gench}) on the finite interval $z \in (-L,L)$ with zero Neumann's conditions at the boundaries $z = \pm L$ and appropriate conditions at $z= \pm d/2$, is applied. At each iteration the corresponding linear boundary value problem is solved numerically using the Finite Elements Method on a nonuniform grid, condensed to the boundaries $z=\pm d/2$ of the layer.

We note that Eqns. (\ref{eq12})-(\ref{eq15}), the boundary conditions at $z=\pm L$, as well as the Veierstrass conditions at the points $z= \pm    d/2$, can be interpreted as a necessary extremum conditions for the free energy functional
\begin{equation}\label{fen}
F[R, \varphi]=\int\limits_{-L}^{L} \Lambda(x, R, R^\prime, \varphi^\prime ) dx + J \left[ \varphi(-L)-\varphi(L) \right].
\end{equation}
\noindent Here, the energy density $\Lambda$ is given by
\[
\Lambda = \frac{1}{2}\left\{ {\begin{array}{*{20}c}
   {R^{\prime 2}  + R^2 \varphi^{\prime 2} - R^2  + \frac{1}{2}R^4 , \>\> \qquad {\rm}z \notin \left({- \frac{d}{2},\frac{d}{2}} \right)}  \\
   {\frac{1}{m}\left( {R^{\prime 2} + R^2 \varphi^{\prime 2} + \alpha R^2  + \frac{\beta }{2}R^4 } \right),z \in \left( { - \frac{d}{2},\frac{d}{2}} \right)}  \\
\end{array}} \right.
\]

All numerical results from now on were obtained for $L=16$ and width of the layer $d=0.2$. Therefore, the two superconducting layers conform to the finite intervals $(-L, -d/2)$ and $(d/2, L)$.
\begin{figure}[ht]
\centerline{\includegraphics[width=3.1in,height=2.45in]{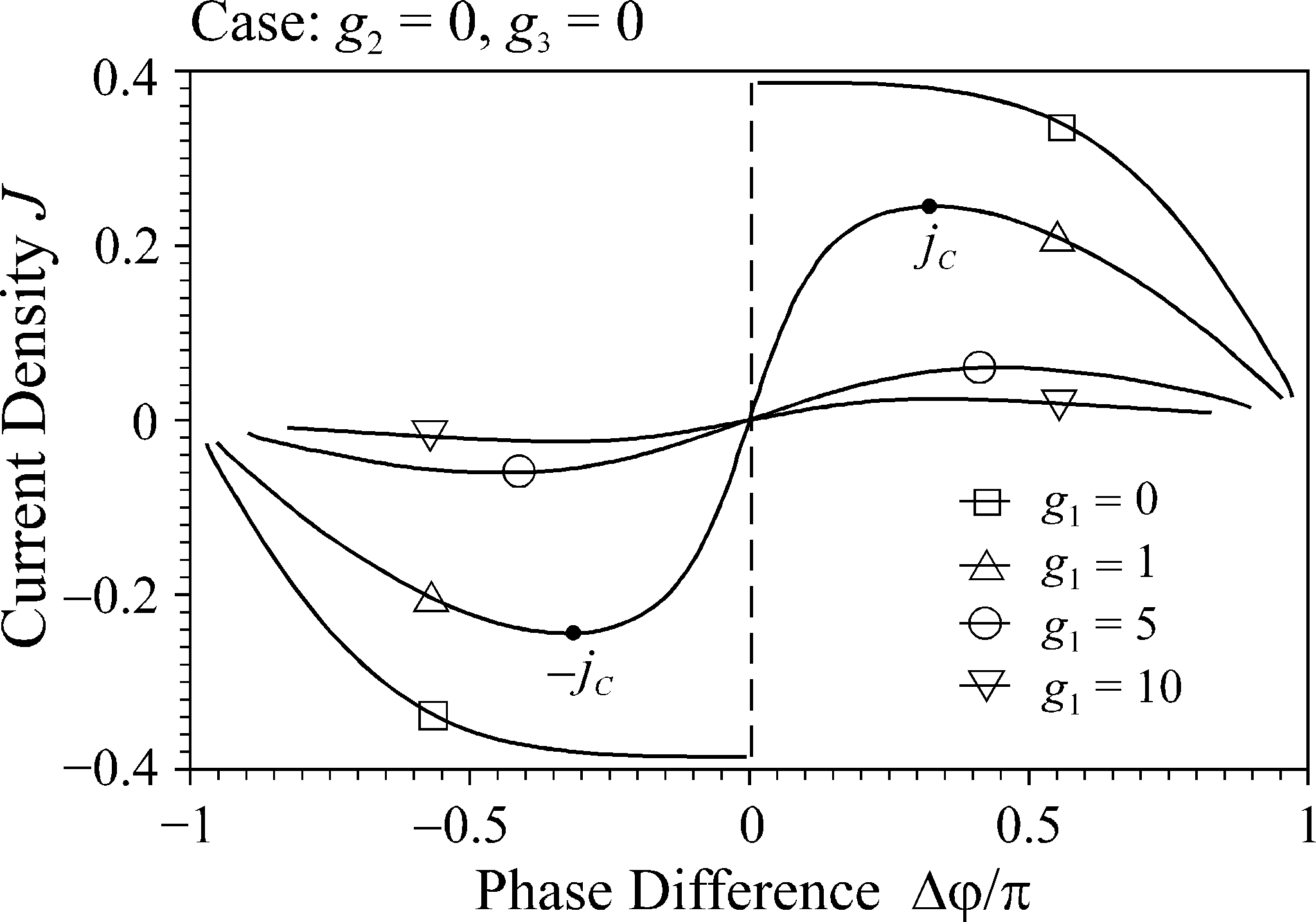}} \caption {Some of curves $J(\Delta\varphi)$ for $g_2=0$ and $g_3=0$.}
\end{figure}
The graphics displayed in Fig. 1 correspond to $J(\Delta\varphi)$ curves for four different values of $g_1$ ($g_1=0, g_1=1, g_1=5$, and $g_1=10$) at $g_2=0$ and $g_3=0$.
\begin{figure}[ht]
\centerline{\includegraphics[width=3.1in,height=2.45in]{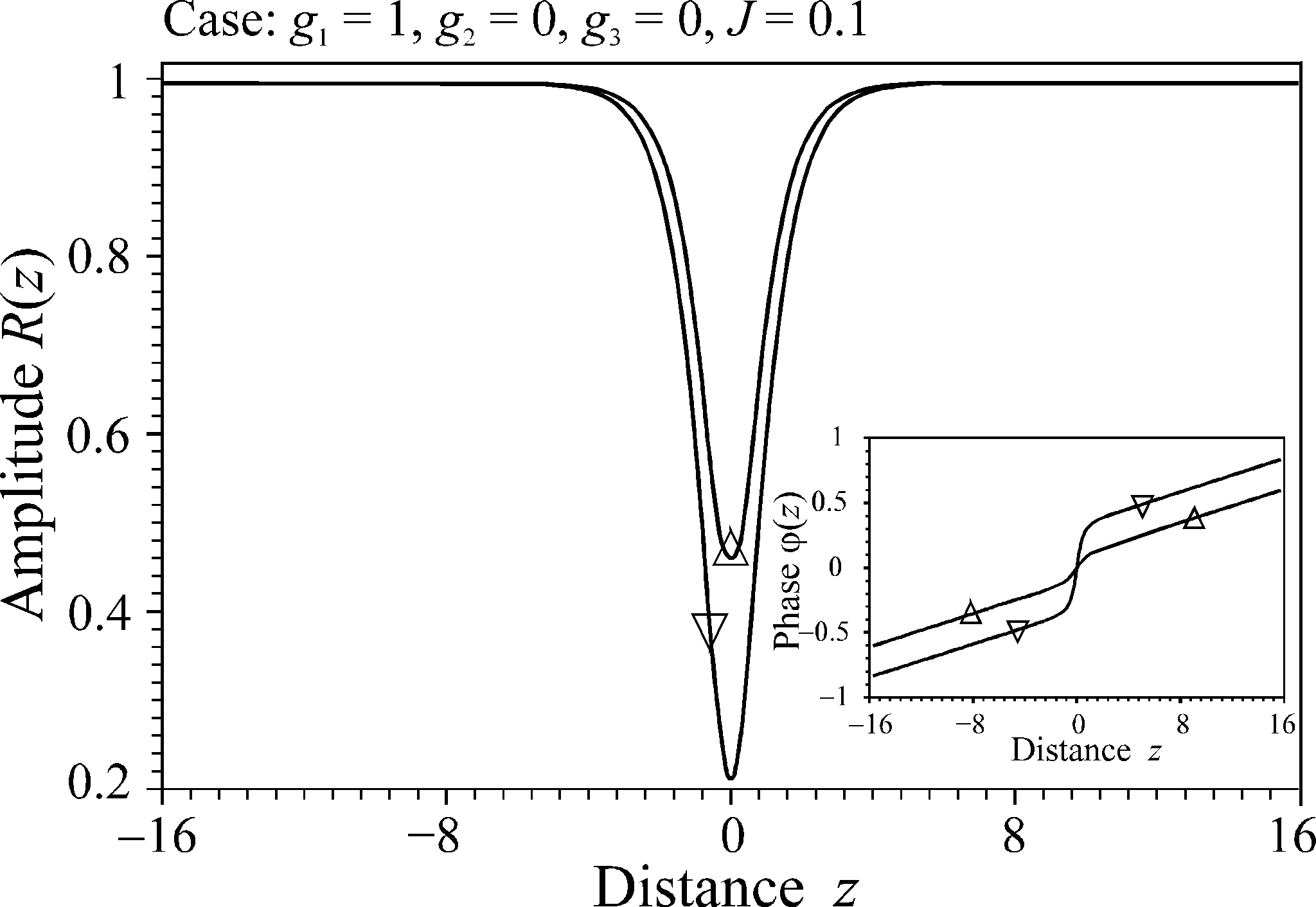}}   \caption {The solutions for $J = 0.1, g_1=1, g_2=0$ and $g_3=0$.}
\end{figure}
If the quantity $g_1=0$ (the corresponding curve is marked by $\square$) the maximum is achieved at $j_{dep} = 2/3 
\sqrt{3} \approx0.385$ (uniform superconductor). For large $g_1$ $(g_1 = 5; g_1 =10)$ we found results close to the ideal Josephson relation $J = j_c \sin \Delta\varphi$, which will be analyzed more strictly below. We note that the numerical results, displayed on Fig. 1 are in good agreement with Fig. 2 in \cite{Sols}. For each curve on this figure we denote $jc = \max J(\Delta \varphi)$ when $\Delta \varphi/\pi \in [-1, 1]$.

For a given value of the current density $J$ we found numerically two solutions, whose amplitudes $R(z)$ and phases $\varphi(z)$ are demonstrated on Fig. 2. For the first (``upper") solution (marked by $\triangledown$) we have the phase offset $\Delta \varphi/\pi \in [-1, -j_c)$ and $\Delta \varphi/\pi \in (j_c, 1]$, while for the second solution (marked by $\triangle$) we have ($\Delta \varphi/\pi \in (-j_c, j_c)$). The first solution originates from the ``uniform" solution $R(z) = 1$, $\varphi(z)=0$, existing in the case when $g_1=0$, $g_2=0$, $g_3=0$, and $J=0$.

The dependence of the free energy $F(J)$ on the current density $J$ for these two solutions is represented graphically on Fig. 3. This is a typical bifurcation diagram: in the point $B$ we have $J=j_c$, the two curves coalesce and acquire a common cusp.
\begin{figure}[ht]
\centerline{\includegraphics[width=3.1in,height=2.45in]{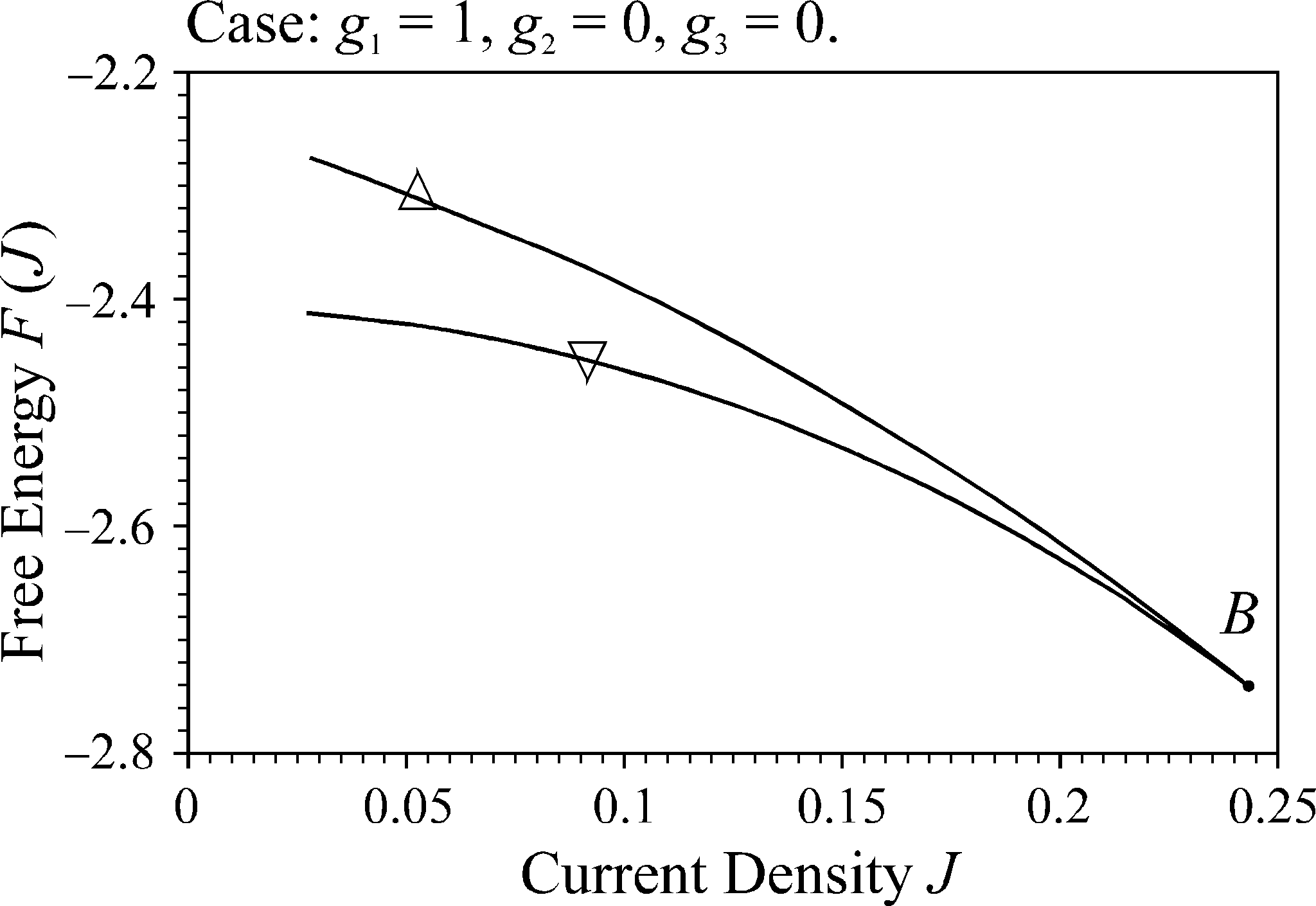}}  \caption {The critical current $j_c$ corresponds to a bifurcation point.}
\end{figure}

\begin{figure}[ht]
\centerline{\includegraphics[width=3.1in,height=2.45in]{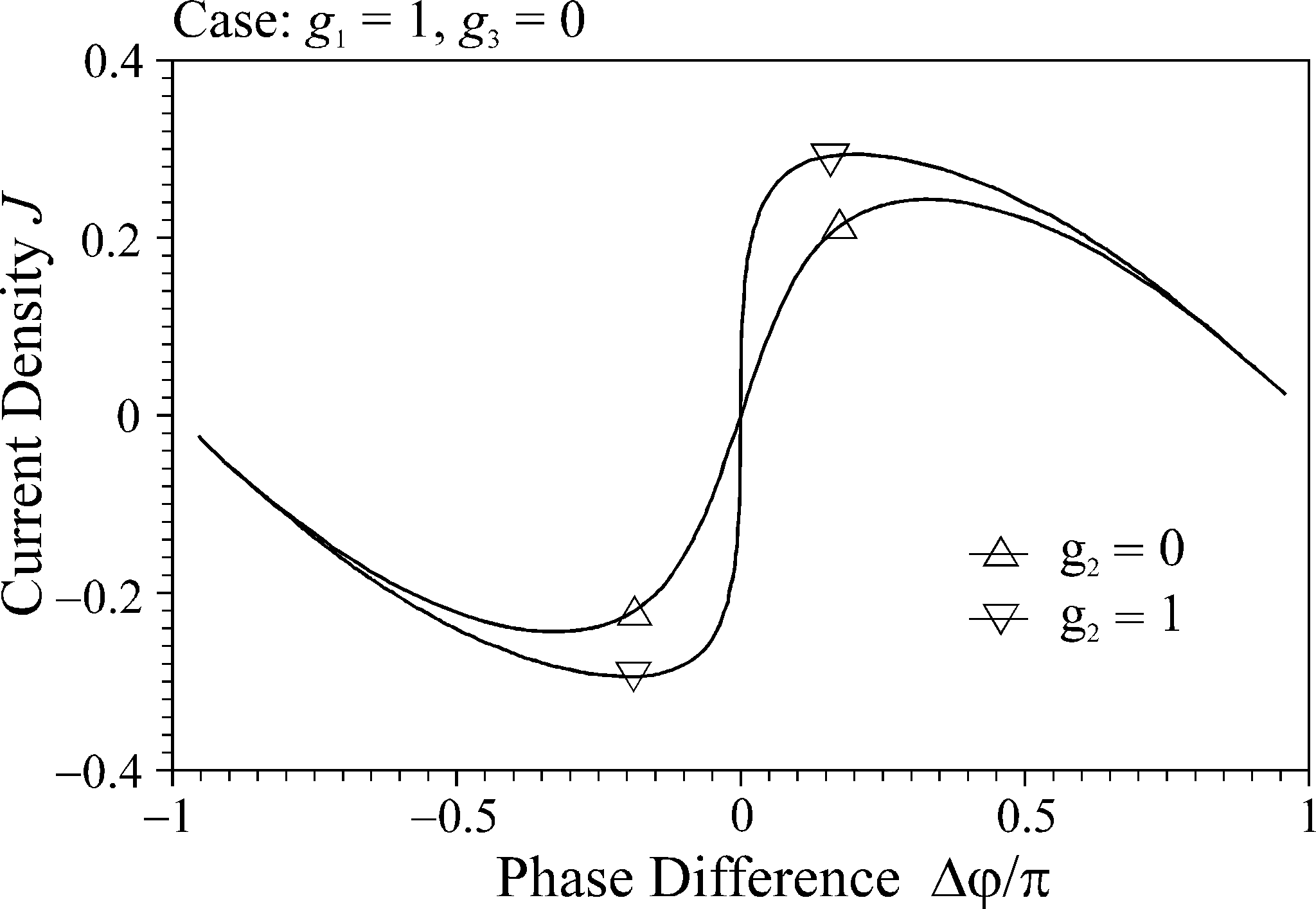}}  \caption {The Influence of the Parameter $g_2$.}
\end{figure}
Figures 4 and 5 are numerical investigations of the influence of the parameters $g_2\neq 0$ and $g_3\neq 0$, respectively, on the $J(\Delta\varphi)$ curve. A comparison between Fig. 4 and Fig. 1 at $g_1=1$ clearly shows that the influence of the parameter $g_2\in [0,1)$ on the current density is not very pronounced (a few percent), whereas a comparison between Fig. 5 and Fig. 1 at $g_1=1$ proves that the variation of the parameter $g_3$ between 0 and -3.5 leads to a significant reduction of the maximum current density (approximately twice).
\begin{figure}[ht]
\centerline{\includegraphics[width=3.1in,height=2.45in]{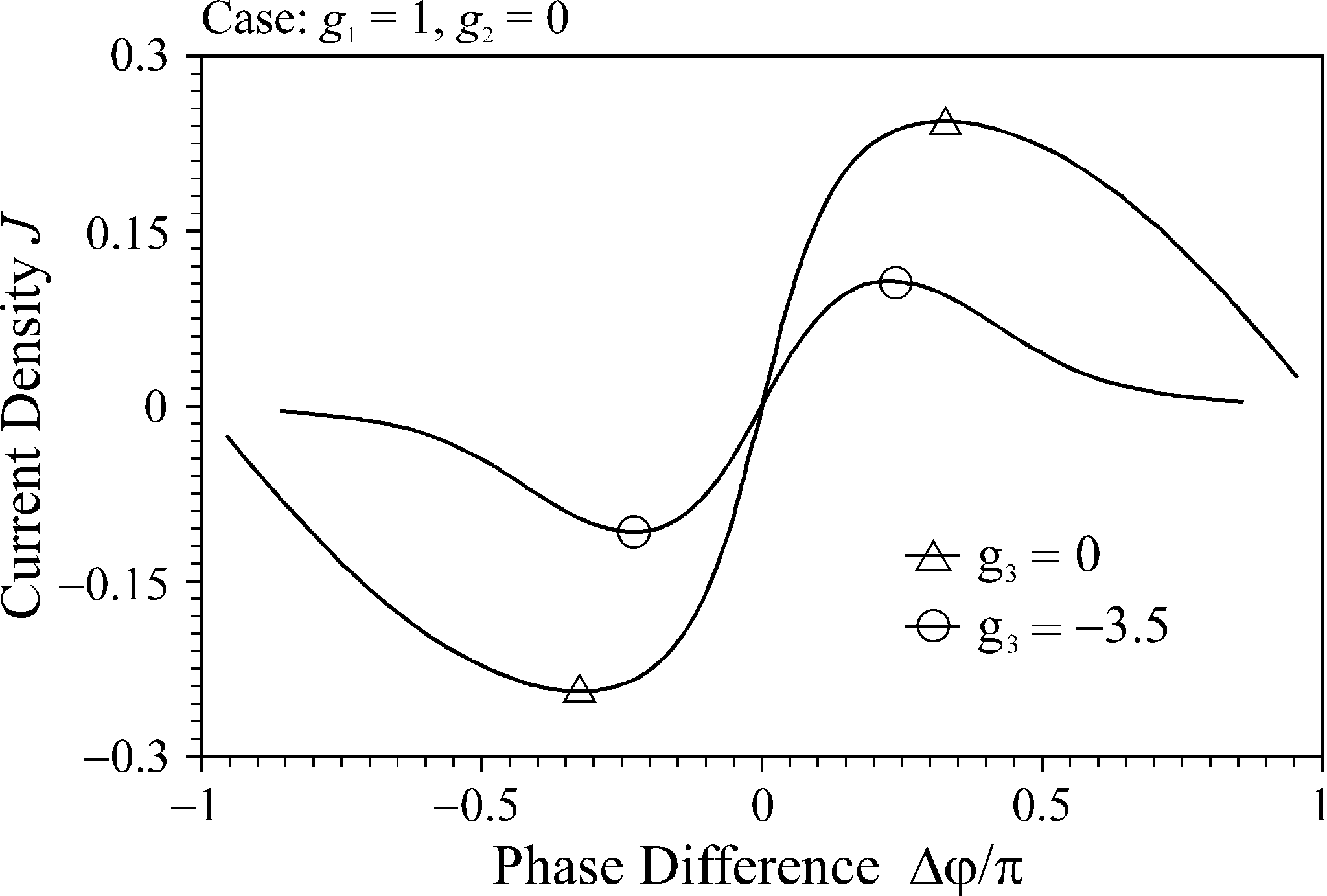}}  \caption {The Influence of the Parameter $g_3$.}
\end{figure}

These quantitative conclusions can be coupled with the Fourier decomposition of $J(\Delta\varphi)$ curves as given by Eqn. (\ref{B}). We restrict ourselves only to the analysis of the ratio $a_2/a_1$ of the first two Fourier coefficients. When $a_2/a_1 \ll 1$ we have an approximately pronounced Josephson behaviour $J\simeq j_c \sin\Delta\varphi = a_1\sin\Delta\varphi$.
\begin{figure}[ht]
\centerline{\includegraphics[width=3.1in,height=2.45in]{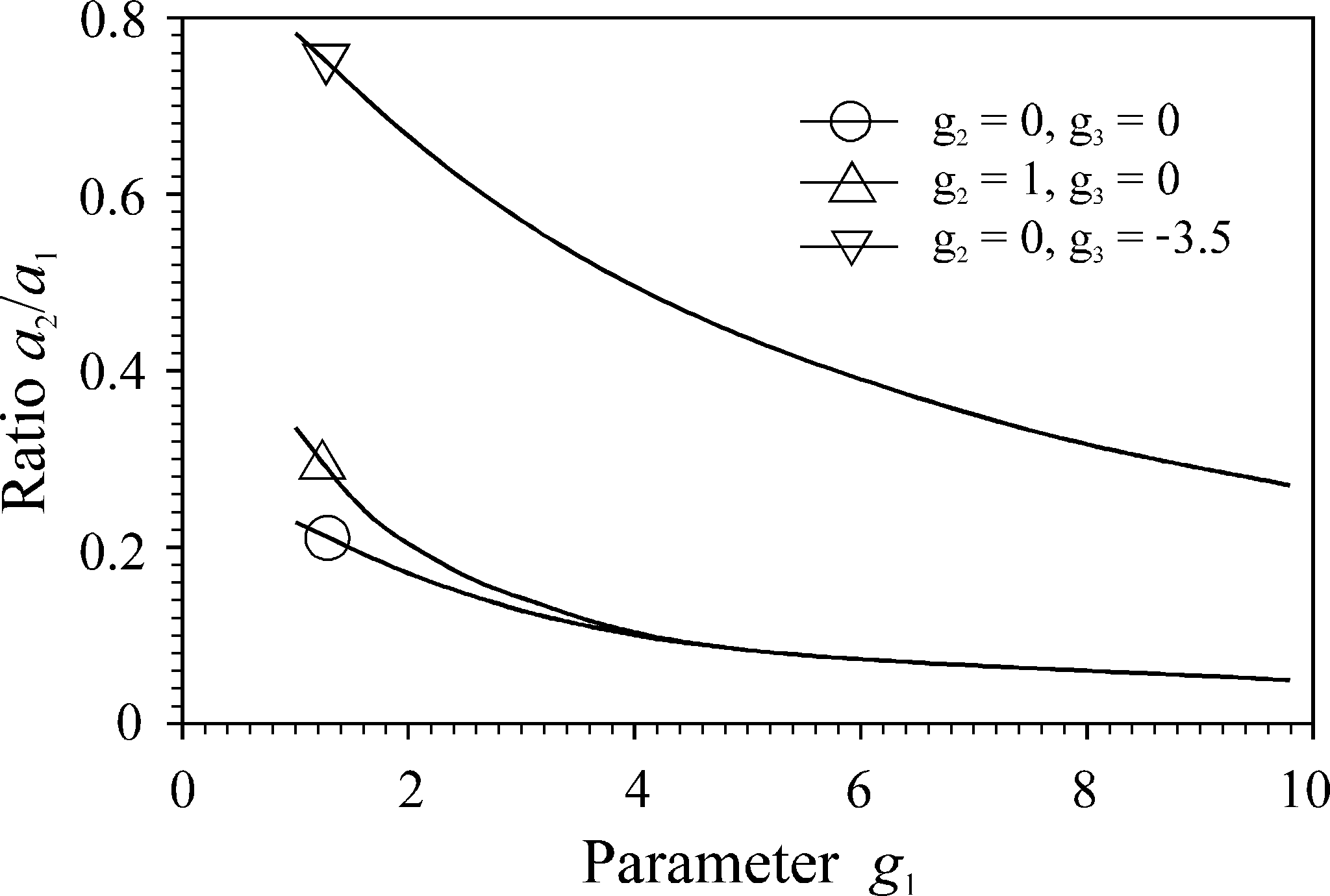}}  \caption {The Ratio $a_2/a_1$ of the first two Fourier coefficients as a function of $g_1$.}
\end{figure}

The ratio $a_2/a_1$ as a function of the parameter $g_1$ is shown on Fig. 6. It is seen that for large values of the parameter $g_1$ ($g_1>8$) when the parameter $g_3=0$, the coefficient $a_2$ is less than $5\%$ of $a_1$. On the contrary, for small values of $g_1$ we have a substantial weight of higher harmonics (for example, if $g_1=1$ then the ratio $a_2/a_1 \simeq 0.23$).

As it can be expected (see the curve marked by $\bigtriangleup$), the influence of the alteration of the parameter $g_2$ on the Fourier coefficients is essential for small enough values only of the parameter $g_1$ ($g_1<4$). On the other hand, taking into account the coefficient $g_3\neq 0$ ($m \neq 1$) leads to a significant increase of the second term in Eqn. (\ref{F1}) (the corresponding curve $a_2/a_1$ is marked by $\triangledown$).

Let us consider the special case of small current $J\rightarrow +0$. In this case, from the usual sinusoidal Josephson relation (\ref{F1}) in linear approximation we get
\begin{equation}\label{f37}
  \Delta\varphi = \arcsin (J/j_c)\simeq J/j_c.
\end{equation}
Eqn. (\ref{f30}) is simplified considerably if $J=0$ and reduces to the following equation
\begin{equation} \label{f38}
  \sqrt{2} (1-Y^2_0)=g_1 Y_0 -g_2 Y^3_0,
\end{equation}
\noindent where $Y_0=y_0 (J=0)$. For small $J$ we have $b\simeq 1, q=2J^2,$ $B=1/ (\sqrt{2}\, J),$ $ 1\gg B^{-1}$ and the right-hand-side of Eqn. (\ref{eq28}) can be replaced by a term, proportional to $J$
\begin{equation}\label{f39}
  \Delta\varphi=2J \left[\sqrt{2} \left(\frac{1}{Y_0}-1\right)-\frac{d(1-m)}{2Y^2_0}\right].
\end{equation}

By coupling Eqns. (\ref{f37}) and (\ref{f39}) we derive an approximate estimation for the critical current
\begin{equation}\label{f40}
  \frac{1}{j_c}=2\sqrt{2}\left(\frac{1}{Y_0}-1\right)-\frac{d(1-m)}{Y^2_0},
\end{equation}
\noindent where $Y_0$ is the smallest root of Eqn. (\ref{f38}).
In the special case $g_2=0, m =1 (g_3=0)$, $g_1\gg \sqrt{2}$, $Y_0 = \sqrt{2}/g_1 \ll 1$, we recover the result from \cite{Sols}
$$ j_c = \frac{Y_0}{2\sqrt{2}}=\frac{1}{2g_1}.$$
\begin{figure}[ht]
\centerline{\includegraphics[width=3.1in,height=2.45in]{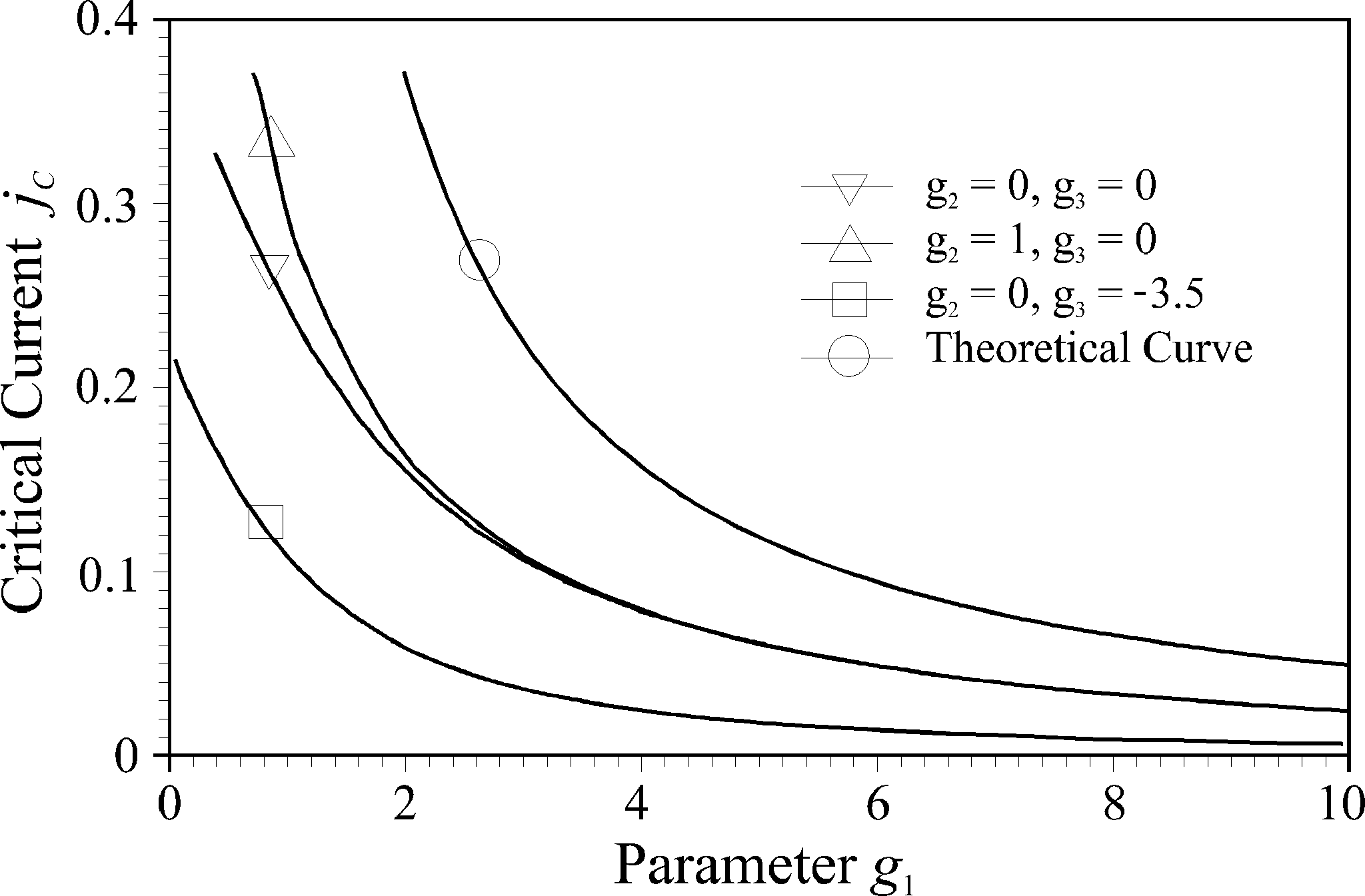}} \caption {The Critical Current $j_c$ as a function of the parameter $g_1$.}
\end{figure}
Figure 7 shows the comparison between numerically obtained and theoretically calculated curves $j_c(g_1)$ by means of formula (\ref{f40}) for different values of parameters. We emphasize the agreement between the theoretical and numerically obtained relations.

\section*{Concluding Remarks}

The physics of Josephson junctions is based on a usual sinusoidal supercurrent-phase difference relation. In the present paper, we show that, by taking into account different nonlinear terms in the normal and superconducting regions, many harmonics exist and the dependence $J(\Delta\varphi)$ of the current as a function of the phase offset is not sinusoidal. This effect follows from the numerical investigation and is seen also from the approximate analytical expressions given above.

We show that the maximum current density $j_c$ represents a bifurcation point for the amplitudes of the supercurrent flow by a change of the current density $J$.

\end{document}